\documentclass[aps,twocolumn,showpacs,superscriptaddress,floatfix,notitlepage]{revtex4-1}
\usepackage[utf8]{inputenc}
\usepackage{braket}
\usepackage{amsmath,amssymb,graphicx}
\usepackage{qcircuit}
\usepackage{color}
\usepackage{appendix}
\usepackage{natbib}
\usepackage{lipsum}
\usepackage[colorlinks=true,citecolor=cyan]{hyperref}
\hypersetup{colorlinks=true,citecolor=cyan,linkcolor=red,urlcolor=magenta}

\date{\today{}}

\begin{document}
\title{Determining ground-state phase diagrams on quantum computers via a generalized application of adiabatic state preparation}

\author{Akhil Francis}
\affiliation{Department of Physics, North Carolina State University, Raleigh, North Carolina 27695, USA}

\author{Ephrata Zelleke}
\affiliation{Department of Physics, Goucher College, Baltimore, Maryland 21204 USA}

\author{Ziyue Zhang}
\affiliation{Department of Physics, Georgetown University, Washington, DC 20057 USA}

\author{Alexander F. Kemper}
\affiliation{Department of Physics, North Carolina State University, Raleigh, North Carolina 27695, USA}

\author{J. K. Freericks}
\affiliation{Department of Physics, Georgetown University, Washington, DC 20057 USA}

\begin{abstract}
Quantum phase transitions materialize as level crossings in the ground-state energy when the parameters of the Hamiltonian are varied. The resulting ground-state phase diagrams are straightforward to determine by exact diagonalization on classical computers, but are challenging on quantum computers because of the accuracy needed and the near degeneracy of competing states close to the level crossings. In this work, we use a local adiabatic ramp for state preparation to allow us to directly compute ground-state phase diagrams on a quantum computer via time evolution. This methodology is illustrated by examining the ground states of the XY model with a magnetic field in the $z$-direction in one dimension.  We are able to calculate an accurate phase diagram on both two and three site systems using IBM quantum machines.
\end{abstract}

\maketitle

\section{Introduction}

Quantum computers are thought to enable calculations that cannot be carried out on classical computers~\cite{lloyd1996universal,abrams1997simulation}. One challenging problem in many-body physics is to determine the zero-temperature phase diagram of finite systems that have level crossings in the ground state as parameters in the Hamiltonian are varied~\cite{qin2021hubbard}. Such phase diagrams commonly occur when a system has competing order parameters~\cite{hubbard1963electron}. One possible approach to solving this problem is to simply create circuits for target wave functions that can have their parameters varied to allow for a variational determination of the approximate ground state. Then, one can determine the phase diagram by examining the quantum numbers and the symmetries of the variational wave function. But, such an approach is likely to fail or to be inaccurate; this is because there are low-lying states near the level crossings and the variational calculations need to be done with high accuracy to carry out such a program. This becomes especially complicated if the variational state ansatz does not belong to the subspace corresponding to ground state quantum numbers.

Another approach one could try is to use adiabatic state preparation: start the system in an easy to prepare state that is the ground state of the Hamiltonian for a given parameter, and then slowly change the parameters in the Hamiltonian. If we change slowly enough, the adiabatic theorem guarantees that we stay in the ground state. This approach may also have problems, because the time evolution will preserve the symmetry of the wavefunction, and level crossings can only occur between states with different symmetries. 

However, we can modify the adiabatic state preparation protocol by adding a small symmetry breaking field, and we can find the phase transition point by monitoring the expectation value of quantum numbers corresponding to different symmetries. Now, because the symmetries are only approximate, a sufficiently slow time evolution will map out the ground-state phase diagram. We then repeat with different magnitudes of the symmetry-breaking field and extrapolate the results to the limit where the symmetry breaking field vanishes. In this fashion, we can employ adiabatic state preparation to carry out a mapping of the ground-state phase diagram. It is unlikely that fast forwarding techniques such as QAOA~\citep{farhi2014quantum} or shortcuts to adiabaticity~\citep{chen2010fast,guery2019shortcuts}, will help with carrying out this approach because it may require very accurate optimization near the level crossing, or knowledge of the eigenstates or invariants of motion which maybe costly to find.

We test our approach on the ground state phase diagram of an isotropic 1D XY model in a magnetic field along the $z$-direction. This system is a stringent test for such an approach, because there are $N$ phase transitions for an N site system in the region $|B_z|\leq |J|$. As the system size is made larger, the problem becomes increasingly more challenging to solve. In fact, the model may exhibit  a devil's staircase in the ground-state phase diagram \citep{bak_one-dimensional_1982}. The conserved symmetry (quantum number) is the $z$-component of total spin, so we can monitor the phase diagram by measuring the magnetization of the system.   

Our strategy is to start the system in a large $B_z$ field, and to add a small symmetry-breaking field $B_x$ in a perpendicular direction. The initial state will be taken to be polarized along the $z$-direction, which is easy to prepare. We ramp the $z$-field down, keeping the $x$-field fixed, using a local adiabatic ramp~\citep{richerme2013quantum}. This approach was originally used to generate the ground state of the transverse-field Ising model in ion trap quantum simulators. For the two site system we also performed the experiment starting from all spins aligned down and ramping up the $B_z$ field. We find that for two and three site system this approach gives accurate phase diagram in IBM quantum machines.
\begin{figure}[b]

    \centering
     \includegraphics[width=0.4\textwidth]{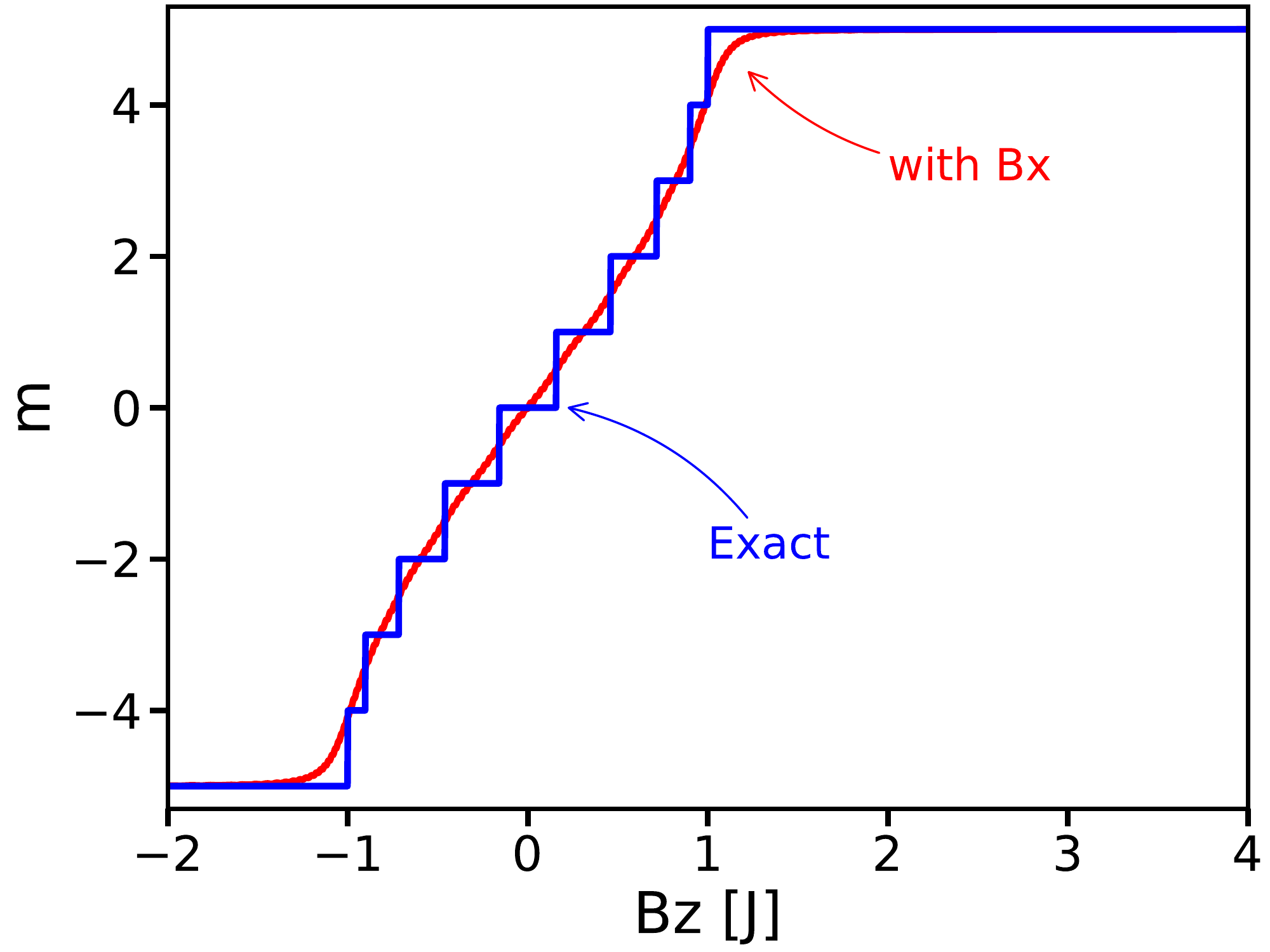}
     \caption{Magnetization versus the $B_z$ field for a 10 site system. The blue curve labeled exact shows the magnetization for the ground state without any $B_x$ field. The red curve shows the magnetization of a local adiabatic evolved state from an all  up state with a local ramp of $\gamma=50$ and 1000 Trotter steps. Here, we have set $B_x=0.05 J$.}
    \label{fig:10site curve}
    
\end{figure}

\section{Methods}

We work with the one-dimensional isotropic XY model with periodic boundary conditions and a magnetic field along the $z$ direction as shown in Eq.~(\ref{eqn:Ham}) for a system with $L$ spins:
\begin{equation}
H = - \sum_{i=1}^{L}\left[\frac{J}{4} \left(\sigma_{i}^{x} \sigma_{i+1}^{x}+ \sigma_{i}^{y} \sigma_{i+1}^{y}\right)+ \frac{B_z}{2} \sigma_{i}^{z}\right].
\label{eqn:Ham}
\end{equation}
where $\sigma^{x,y,z}_i $ are the usual Pauli matrices obtained by setting $(\hbar =1)$ in the spin operators of the $i^{th}$ site, $S^{x,y,z}_i = \frac{\hbar}{2}\sigma^{x,y,z}_i$. This model can be solved exactly by fermionization using a Jordan Wigner (JW) transformation~\cite{jordan1928pauli} and a subsequent Fourier transformation to work in momentum space~\cite{lieb1961two}. The boundary term needs more care as it still has the Jordan Wigner string in it. Usually for a large system this term is negligible. Alternatively, we simply consider the periodic term without a JW string attached, so that the usual Fourier transform yields the fermionic eigen energies. Then the fermionic Hamiltonian takes the form
\begin{equation}
H=\sum_{k}\left( \omega_{k} c_{k}^{\dagger} c_{k} -\frac{B_{z}}{2}\right)
\end{equation}
where 
$
\omega_{k}=\left(-J \cos k + B_{z} \right)
$
and $k=\frac{2 n \pi}{L}$, with $n=-\frac{L}{2}+1, \cdots, 0, \cdots, \frac{L}{2}$. For a finite-size system the boundary term matters; this can be dealt with by making use of the fermionic parity \citep{mbeng2020quantum}. We work in original spin representation throughout this paper.

\begin{figure}[htpb]
    \centering
    \includegraphics[ width=0.4\textwidth]{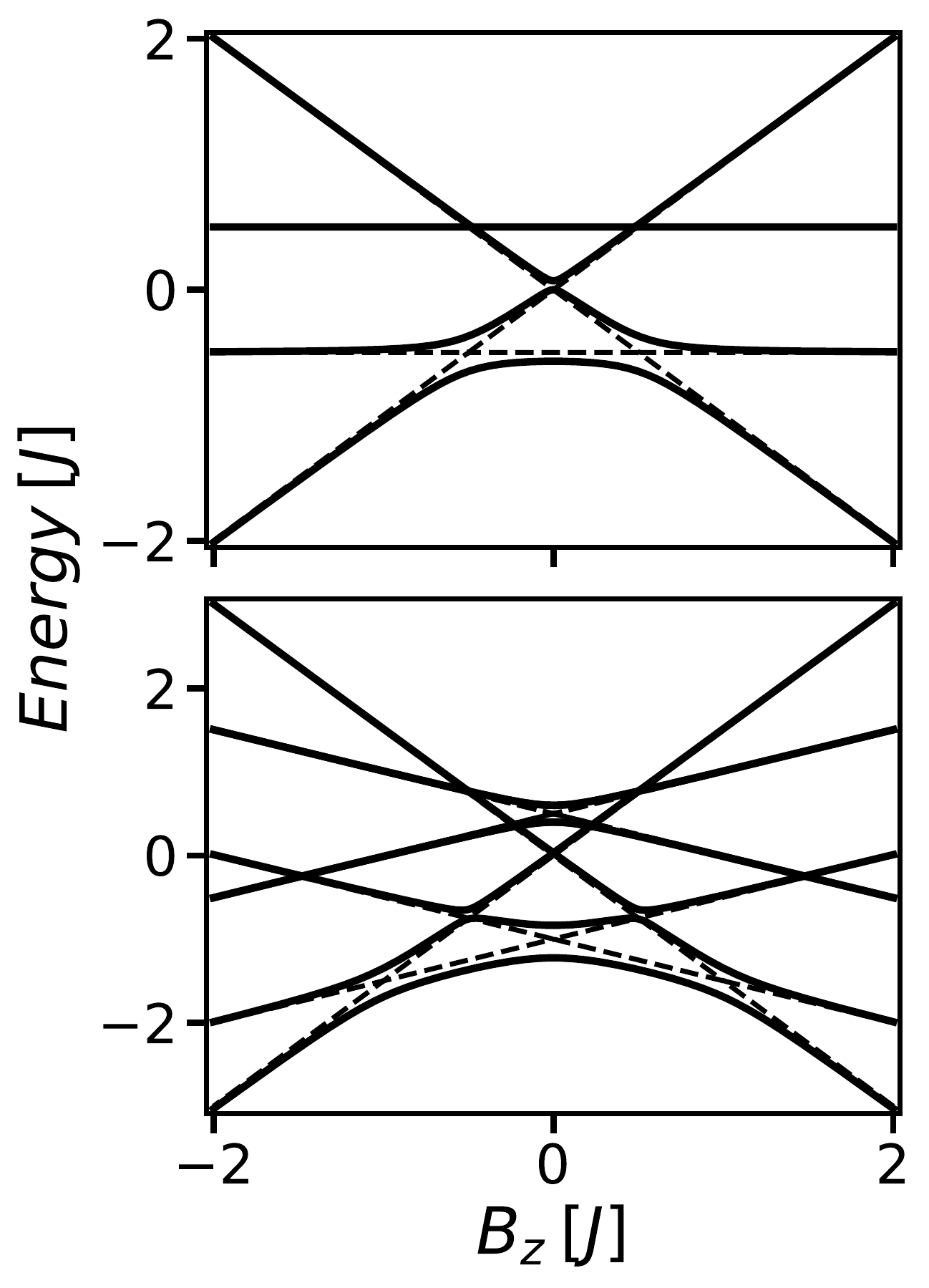}
    \caption{Energy diagram for a two site (top panel) and three site (bottom panel) system. The dotted lines are for the ideal system with $B_x=0$; these curves have level crossings. The solid lines are the energies with $B_x=0.2 J$, which leads to avoided crossings everywhere for the ground state.}
    \label{fig:energy diagram}
\end{figure}

\begin{figure}[htpb]
    \centering
\includegraphics[ width=0.4\textwidth]{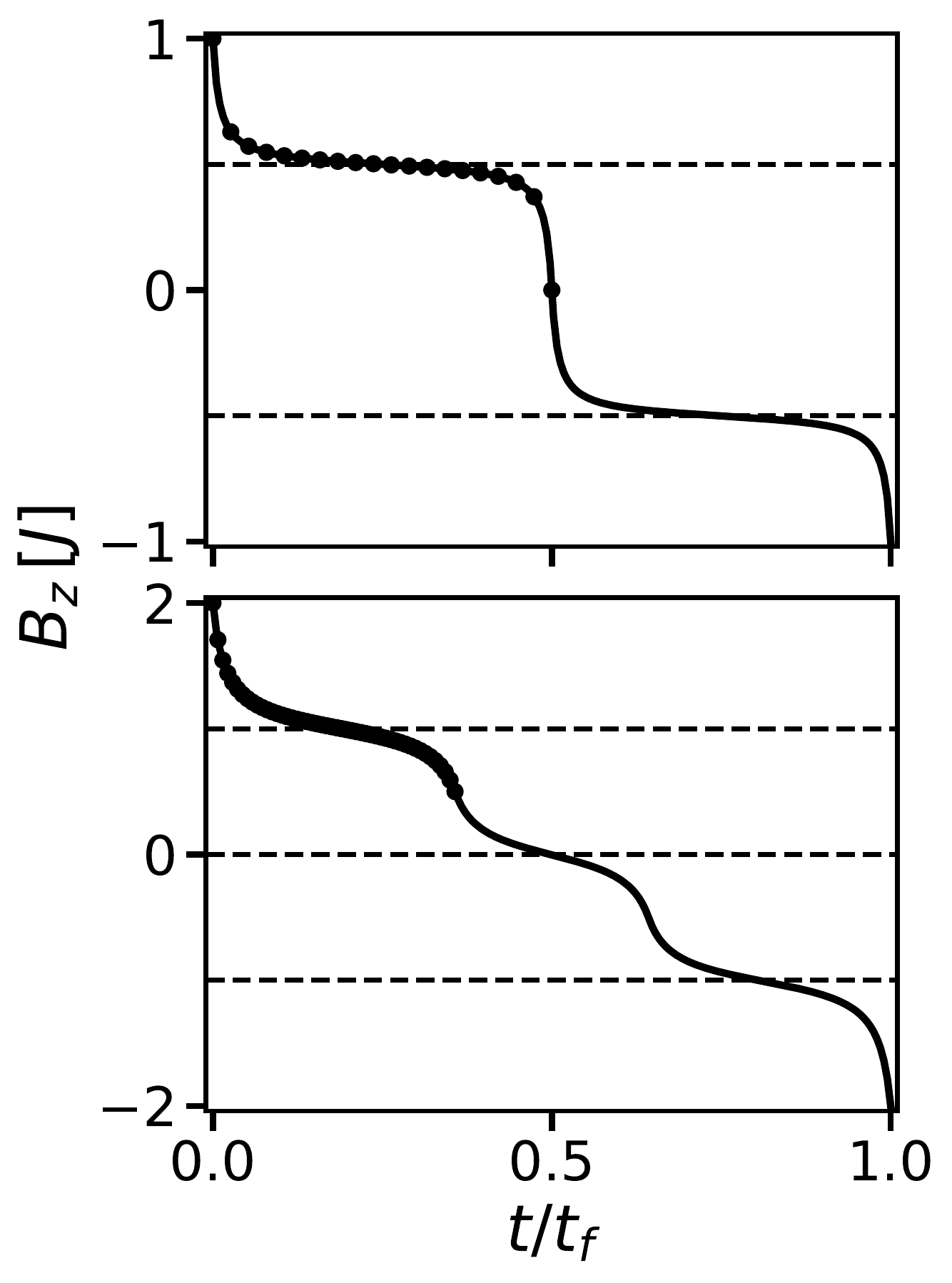},
    \caption{Local adiabatic ramp for a two site (top panel) and three site (bottom panel) system. For the two site system, the solid line shows a ramp with 200 time steps and the solid circles are for 20 time steps. Here, we have set $B_x=0.02J$ and $\gamma = 1.5$ for two sites. For the three site system, we have taken $B_x=0.08J$ and $\gamma = 2$ and the curves are similar (solid line, 200 time steps; solid circles, 50 time steps). The horizontal axis is the fractional time $t/t_f$ for the 200 step ramp. We use an Akima-spline interpolator to obtain the ramp in uniform time steps from uniform magnetic field steps used in evaluating the integral. One can see that the ramp goes slower near the crossing regions.}
    \label{fig:local ramp}
\end{figure}

The ground state has many level crossings as a function of the magnetic field $B_z$. This is illustrated in Figs.~\ref{fig:10site curve} and~\ref{fig:energy diagram}. 
Fig.~\ref{fig:10site curve} shows the expectation value of the $z$-component of spin (also known as the magnetization) as a function of $B_z$. Each of the vertical steps on the exact curve corresponds to a level crossing, where the quantum-number for the $z$-component of spin shifts by one unit; the plot also shows an adiabatic time evolution, which will be discussed later.
In this work, we show how to obtain these quantum phase transition points (critical $B_z$ values) on a quantum computer. 

In adiabatic state preparation, we start from the ground state of a Hamiltonian which is easy to prepare and then we slowly evolve the state using time evolution with a Hamiltonian that interpolates from the initial Hamiltonian to the target Hamiltonian. The amount of diabatic excitations are determined by how fast the Hamiltonian changes near the avoided crossing spectral gaps to higher excited states.  The initial Hamiltonian can be thought of as a Hamiltonian with $J=0$, or equivalently with $B_z\gg J$. Then, the magnetic field is ramped down to let's say zero, and where ideally we end up in the ground state of the Hamiltonian with $B_z=0$. But, this cannot occur if there is additional symmetry in the Hamiltonian. Here, because $S^z=\sum_i \sigma_i^z/2$ commutes with $H$, we can simultaneously diagonalize both operators and this means the quantum number corresponding to the total $z$-component of spin~(m) are unchanged during time evolution. Thus, we only stay in the ground state of a system with definite $z$-component of spin. 
This can be seen in Fig.~\ref{fig:energy diagram} where the dotted lines showing level crossing for a two and three site system.

In order to achieve adiabatic state preparation, we must break the symmetry. We do so by adding a small $B_x$ field, upon which $[S_z,H]\ne 0$. This means states that used to have different $m$ quantum numbers are now coupled together. This can be seen in Fig.~\ref{fig:energy diagram} where the solid lines showing avoided level crossing for a two and three site system with a $B_x$ field. This then allows adiabatic state preparation to take place, and if we go slow enough, we will have limited diabatic excitation out of the ground state. Fig.~\ref{fig:10site curve} shows that with a $B_x$ field one can traverse through all the magnetization sectors in a ten site system. However, the $B_x$ term changes the Hamiltonian and its energy levels. We only have quantum phase transitions when $B_x=0$, which implies we must extrapolate to the $B_x\to 0$ limit.  

The time evolution is implemented with a local adiabatic ramp~\cite{roland2002quantum,  richerme_experimental_2013}, which is supposed to yield the same diabatic excitation for each time step of the time evolution. It does so by ramping faster when the gap to the first excited state is large and more slowly when the gap is small. It is constructed by adjusting the rate $\frac{d B_z}{dt}$ according to the instantaneous energy gap ${\Delta (B_z)}$ such that  $|\frac{d B_z}{dt}| \ll {\Delta ^2 (B_z)}$ \cite{richerme_experimental_2013}. This provides the highest fidelity ramp for a given total time of evolution.

The total time for the local ramp is determined by an adiabaticity parameter ($\gamma$). We require $\gamma\gg 1$ for an adiabatic ramp, where $\gamma=\left|\frac{\Delta^{2}(B_z) dt}{d{B_z}(t)}\right|$. Starting from a magnetic field $B_{z,\mathrm{initial}}$ and ramping down to a magnetic field $B_{z,\mathrm{final}}$ the ramp time t is given by
\begin{equation}
t = - \gamma \int_{B_{z,\mathrm{initial}}}^{B_{z,\mathrm{final}}} \frac{d B_z}{\Delta^{2}(B_z)}.
\label{eqn:ramp}
\end{equation}
where $\Delta$ is the energy difference between first excited state and ground state and the minus sign indicates that we are ramping down. One can either choose the adiabaticity parameter first, and determine the total time, or one can fix the total time and infer the adiabaticity parameter. A resulting local adiabatic ramp for a two and three site system is shown in Fig.~\ref{fig:local ramp}.
The ramp was obtained via a Trotter product formula. We select the adiabaticity parameter and the number of Trotter steps such that we can accurately determine the different steps in the magnetization, which signal the different regions of the ground-state phase diagram. 

The same strategy is used on a quantum computer. We use the Trotter product formula for the time-evolution operator from $t_0$ to $t$: 
\begin{equation}
\mathbf{U}\left(t, t_{0}\right) \approx e^{-i \mathbf{H}(t- dt) dt} e^{-i \mathbf{H}(t-2d t) d t} \ldots e^{-i \mathbf{H} \left(t_{0}\right) dt}.
\end{equation}
Then each Trotter step further is decomposed into two qubit and single qubit gates so that it can be implemented on the IBM machines using their native gate set.

The determination of the phase diagram then proceeds as follows: (i) we initialize the system in a state that is all up and with the magnetic field equal to $B_{zinitial}$ and with a fixed value for $B_x$;
(ii) we evolve the system from $t_0$ to $t$ using the local adiabatic ramp for $B_z(t)$;
(iii) at each time step, we measure the magnetization;
(iv) using the magnetization, we determine the critical value of $B_z$, which corresponds to the midpoint of the step in the magnetization between two successive $m$ quantum numbers;
(v) we repeat these steps for a different value of $B_x$; and
(vi) we extrapolate the critical $B_z$ field to the limit where $B_x\to 0$.

Extrapolation to find critical $B_z$ can be done by fitting polynomial curves to the data. The $\sigma^x$ operator only connects states with definite $m$ eigenvalues that are shifted by one: $n=m\pm 1$, for $\langle n|\sum_i\sigma^x_i|m\rangle\ne 0$. Then, a simple argument using perturbation theory shows that the perturbed energy eigenvalues are functions of even powers of $B_x$ for small $B_x$ values. This means that when we try to extrapolate the exact results for the critical $B_z$ field, we should use a dependence on even powers of $B_x$ only. Hence, we include only quadratic and quartic terms in the fitting curve for the data generated on a classical computer via exact diagonalization. 

For the data from a quantum computer, we instead fit with a linear regression, because the noise on the quantum computer changes the behavior from even powers of $B_x$ to a nearly linear dependence.  When quantum computers become capable of doing longer time runs, with less noise and decoherence, then we can fit a quadratic polynomial without the linear term for smaller $B_x$ values to estimate the critical point in a more systematic way.

\section{Results}

\begin{figure*}[htpb]
    \centering
 \includegraphics[width=0.89\textwidth]{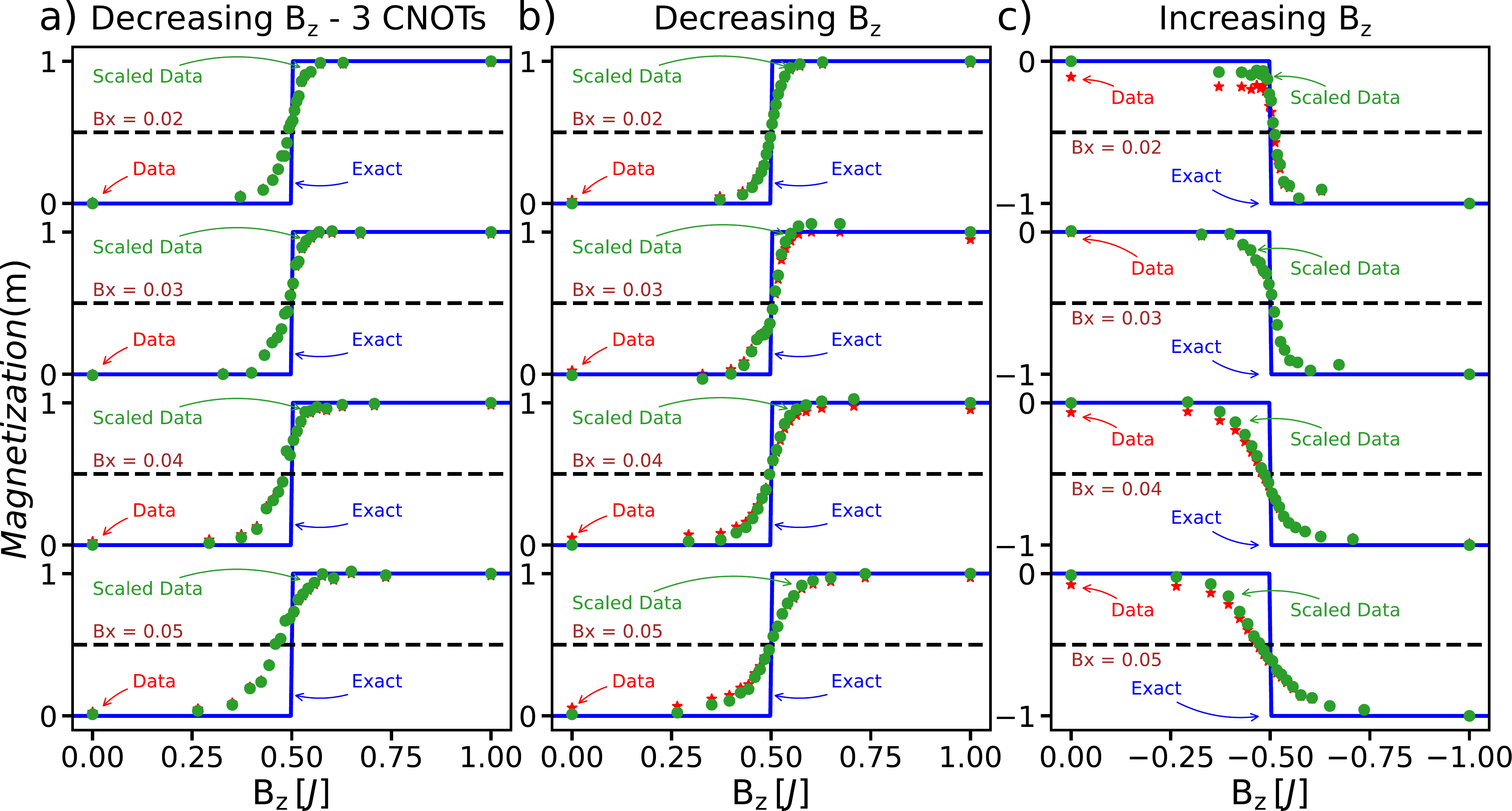}

    \caption{Magnetization versus $B_z$ data for a two-site system. We used a local adiabatic ramp of 20 time steps with $\gamma$ =1.5, starting from the all up state (left, middle) and all down state (right). Exact denotes the magnetization curve with $B_x=0$. Data denotes the values obtained from the IBM Santiago machine after readout error correction. Scaled data modifies the error-corrected data so as to match the known end points. The leftmost panel uses the optimal two-qubit circuit with three CNOTs for the positive magnetization sector. The middle and rightmost panel uses the trotterised circuit for the positive and negative magnetization sector respectively.}
    \label{fig:2 site Exp all up data}
    
\end{figure*}

In order to demonstrate the technique we first examine a two-site system. At large $B_z$, the ground state has all spins aligned in the up direction; the calculation starts with this state. The system is time evolved using the Trotter product formula using a local adiabatic ramp given in Eq.~\ref{eqn:ramp}. The integral produces $t(B_z)$ with uniform steps in $B_z$. We convert to $B_z(t)$ with uniform steps in $t$ by inverting the map and employing an Akima spline, which preserves the shape (see Fig.~\ref{fig:local ramp}). We choose $\gamma$ and the number of time steps such that the time evolution spans the change in magnetization by one full unit (see Fig.~\ref{fig:2 site Exp all up data}). We repeat the same procedure to find the time evolution for each $B_x$ value. The time evolution is implemented in the quantum simulation using two qubit and single qubit gates. We decompose each Trotter step into the XY part, the $B_x$ part and the $B_z$ part:
\begin{figure}[htpb]
\centering
\includegraphics[width=0.35\textwidth]{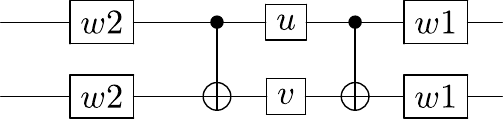}
\caption{Quantum circuit for the evaluation of $\exp\left(-i \mathcal{H}^{XY}_{12} dt\right)$. This is the time evolution circuit for the XY part in each Trotter step for the two site system. $w_{1} = \frac{{\mathbf{1}}-i \sigma^{x}}{\sqrt{2}}$, $w_{2}=w_{1}^{+}$, $u=\exp{\left(i \frac{J}{4} \sigma^{x} dt \right)}$, $v=\exp{\left( i \frac{J}{4} \sigma^{z} dt \right)}$.}
\label{fig:circuit_2site xy}
\end{figure}

\begin{widetext}
\begin{align}
\mathbf{U}_{12}\left(t+dt, t\right) &=
\exp{\left[-i [\frac{J}{4} \left(\sigma_{1}^{x} \sigma_{2}^{x}+ \sigma_{1}^{y} \sigma_{2}^{y}\right)+ \frac{1}{2}B_x \left(\sigma_{1}^{x} + \sigma_{2}^{x}\right) + \frac{1}{2}B_z(t) \left(\sigma_{1}^{z}+\sigma_{2}^{z} \right) ] dt\right]}
\\
& \approx \exp{[-i \frac{B_z(t)}{2} \left(\sigma_{1}^{z}+\sigma_{2}^{z} \right) dt]}
\exp{[-i \frac{B_x}{2} \left(\sigma_{1}^{x} + \sigma_{2}^{x}\right) dt] }
\exp{[-i \frac{J}{4} \left(\sigma_{1}^{x} \sigma_{2}^{x}+ \sigma_{1}^{y}\sigma_{2}^{y}\right) dt]}
\end{align}
\end{widetext}
The XY part is further decomposed to implement in the quantum simulation using two CNOTs~\citep{vidal_universal_2004} as shown in Fig.~\ref{fig:circuit_2site xy}. The $B_x$ part and $B_z$ part are implemented using single qubit gates.

\begin{figure}[htpb]

    \centering
    \includegraphics[width=0.4\textwidth]{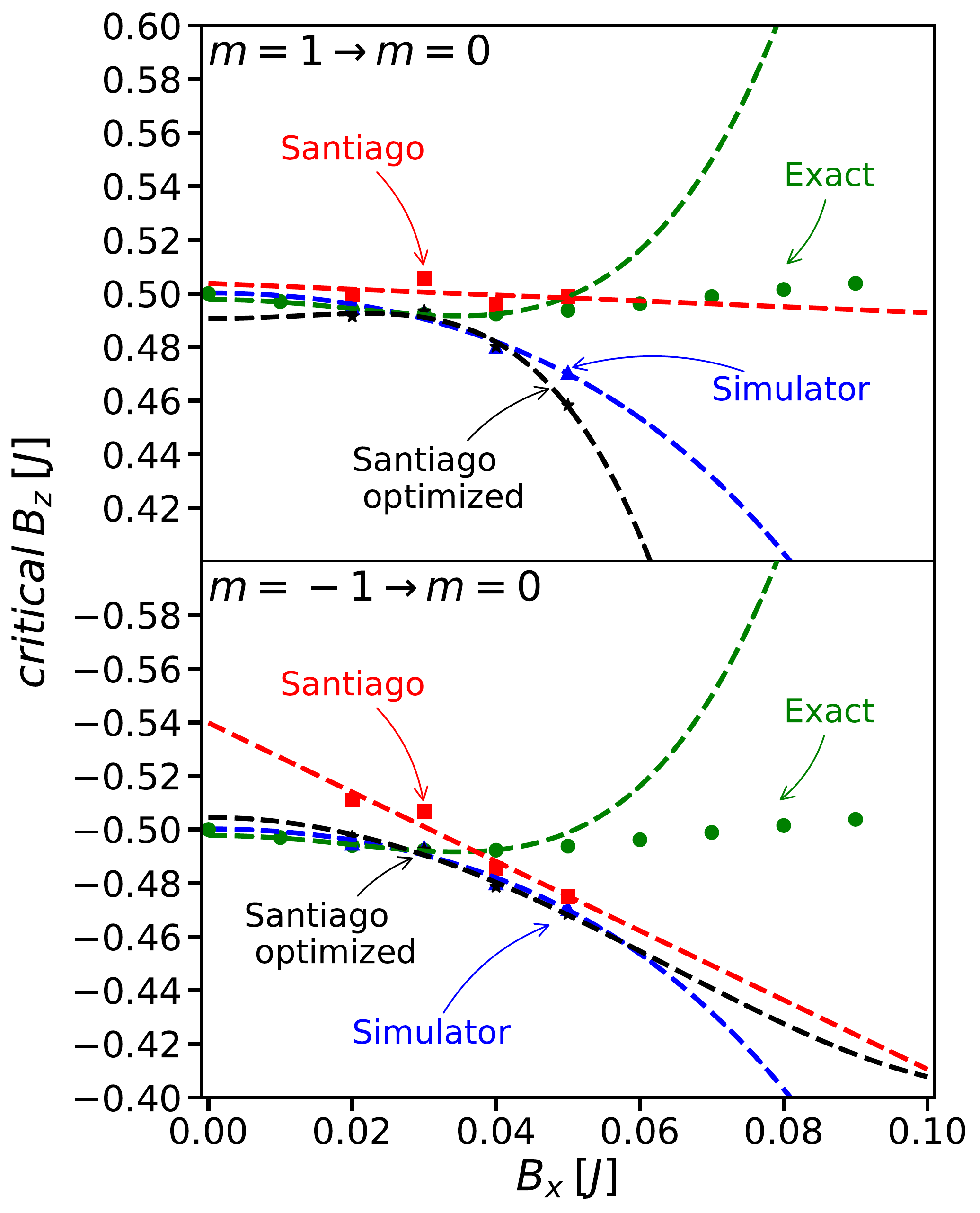}
  
    \caption{Estimation of the quantum phase transition for a two-site model using extrapolation to the $B_x\to 0$ limit. To obtain the critical value for each $B_x$, we interpolate the data points (scaled) and determine where the magnetization is equal to 0.5. For the exact simulation (green solid circles), we have used local adiabatic ramp with 1000 time steps and $\gamma$ value 1.5. The simulator (blue triangles) and quantum computer (red solid squares for circuit without optimization and black stars for circuit with optimization) use 20 time steps. The exact value at $B_x=0.0$ is shown explicitly (green solid circles) to indicate where the actual phase transition is. The initial state has all up spins for the top panel and all down spins for the bottom panel. The green dotted line shows the quartic fit of the exact results to the first four data points. The blue dotted line is the quartic fit to the simulator values. The red dotted line is the linear fit for the quantum computer calculation with no circuit optimization. Black dotted line is the quartic fit for optimized circuit data with three CNOTs.}
    \label{fig:2 site Extraploation to find transition}
\end{figure}

For the two site case, we decrease $B_z$ from $B_z$ = 1.0J to $B_z$ = 0.0J to go through the first transition point (we have set $J = 1$ in all our calculations). 
We use $B_x$ values given by 0.02J, 0.03J, 0.04J, and 0.05J (see Fig.~\ref{fig:2 site Exp all up data}). For the exact curve we use 1000 Trotter steps, but since we cannot achieve high fidelity in currently available quantum quantum computers for such large number of Trotter steps, we look for a similar trend in the crossing point so that a fewer number of Trotter time steps is sufficient (see Fig.~\ref{fig:2 site Extraploation to find transition}). For a two site system the time evolution could be implemented with three CNOTS\cite{vidal_universal_2004} to achieve high fidelity. But since these short depth circuits are not available in general for large system size, we consider explicit implementation of trotter circuit to enable comparison with large system size. Later we also show results from a three CNOT version of the circuit. With 20 Trotter steps the simulator data showed reasonable results. We use an Akima spline to fit the magnetization versus $B_z$ data to have a smooth curve allowing us to determine the transition point. From the simulator data we find the crossing points where the magnetization is equal to 0.5. A quartic fit was performed to the crossing points from the simulator data and we obtain a critical value of $B_z=0.500J$. This result is reasonable, since performing a quartic fit to the first four data points in the exact curve yields an extrapolated value equal to 0.498J.

\begin{figure*}[htpb]
     \centering
     
      \includegraphics[width=0.89\textwidth]{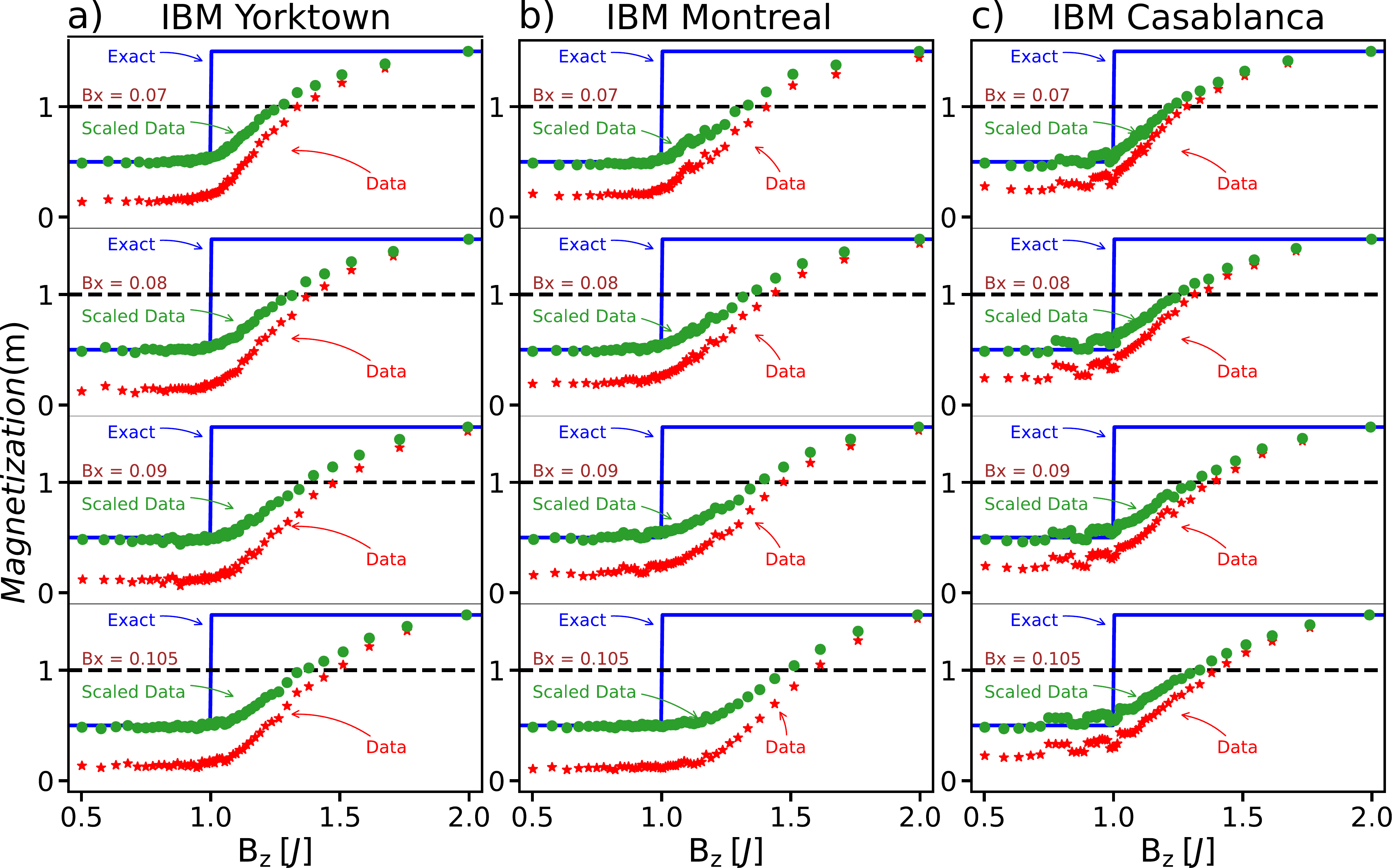}
  
    \caption{Experimental data (after readout error correction) for the magnetization (positive) versus $B_z$ on a three-site system with a local ramp starting from the ground state of all spins up on different machines; a) IBM Yorktown b) IBM Montreal c) IBM Casablanca. We used 50 time steps and $\gamma$ = 2 for the ramp.}
    \label{fig:3 site Exp data}
\end{figure*}
We perform the quantum computer run on the IBM Santiago machine.
The data obtained from the quantum computer with readout error correction is shown in Fig~\ref{fig:2 site Exp all up data}. As a secondary error mitigation technique, we scale the data to match the known initial and final data points to the quantum computer data. This stretches and shifts the data so that the end points have the correct values. Such scaling is common to correct from decoherence and noise, and it improves the data analysis. The experiment is repeated for different $B_x$ values and the corresponding crossing points are plotted in the Fig.~\ref{fig:2 site Extraploation to find transition}. Since the data obtained was noisy we fitted a linear extrapolation curve to capture the trend in the quantum computer data. The extrapolated critical $B_z$ value is 0.504J, which is close to the actual value of transition, which occurs at 0.5J. 
\begin{figure}[htpb]
    \centering
    \includegraphics[width=0.49\textwidth]{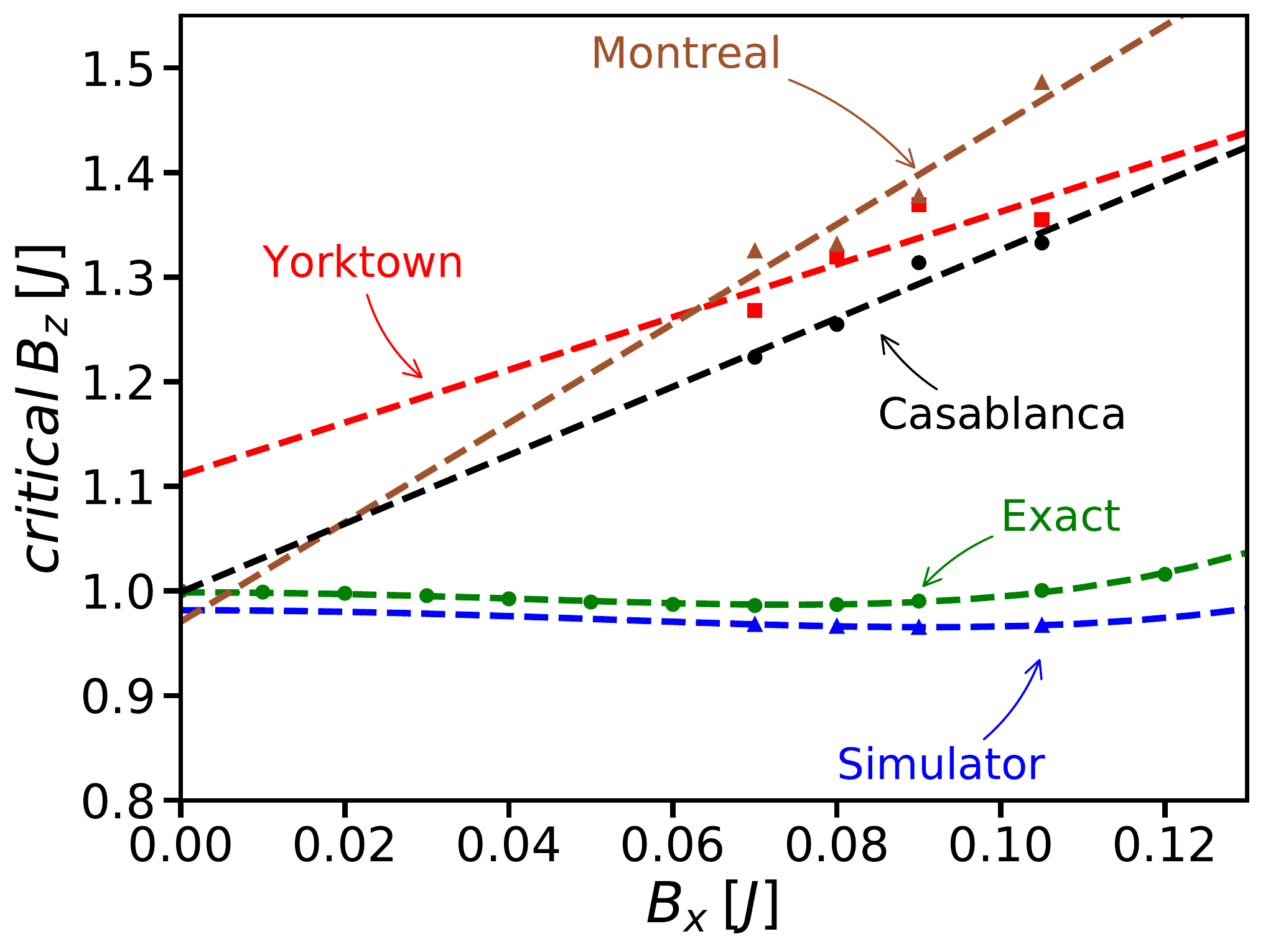}
    \caption{Estimation of the quantum phase transition for three sites using extrapolation to the $B_x\to 0$ limit. This case is for an initial state with all spins up.  To obtain the critical value for each $B_x$, we use an interpolation of the scaled data points. Red squares corresponds to data from IBM Yorktown, brown triangles for IBM Montreal and black circles for IBM Casablanca. The corresponding colored dotted lines shows the linear fit for these data points. For exact calculation (green circle), we use 1000 time steps and $\gamma=2$. For the  simulator (blue triangle) and the quantum computer, we use 50 steps. The exact value at $B_x=0.0$ is shown explicitly (green circle) to indicate where the phase transition is actually located. The quartic fit to the exact and simulator data points is shown with the green and blue dotted lines respectively.}
    \label{fig:3 site Extraploation to find transition}
\end{figure}

For the second case, we examine the transition from $m=-1$ to $m=0$. While, formally, this should be the same as the case for $m=1$ to $m=0$, because the $|1\rangle$ state of a quantum computer is the excited state, decoherence effects should be larger for this case.  Here, the initial state has both spins down. The procedure is similar to what we explained above. A quartic fit to the simulator values gave the critical value to be -0.50 where as a quartic fit to the first four data points of the exact crossings (for $\gamma$ =1.5 and 1000 Trotter steps) gave the extrapolated value to be $-0.498J$. The data obtained from IBM quantum computer is shown in Fig.~\ref{fig:2 site Exp all up data}. The crossing points were found from the scaled data for each $B_x$ values.  A linear fit to the the quantum computer data gave the critical value to be $-0.54J$.

For the two site model, we also performed the experiment in IBM Santiago machine after optimising the circuits with a level three optimization in IBM Qiskit transpiler. This reduced the number of CNOT gates to three for each time. This is because any two qubit unitary operation can be represented using three CNOT gates~\citep{kraus_optimal_2001,vidal_universal_2004}. The read-out corrected data is shown in Fig.~\ref{fig:2 site Exp all up data}. The crossing points from these fixed depth circuits are shown in Fig.~\ref{fig:2 site Extraploation to find transition}. These values are more closer to the simulator values than the trotter data as expected. A quartic fit to these data give the critical $B_z$ value to be 0.491J for all spins aligned up case and -0.504J for all spins aligned down situation. This kind of an efficient fixed depth decomposition is not known in general for more than two qubits. But for certain models efficient fixed depth circuit decompositions can be found~\citep{kokcu2021fixed}. These type of fixed depth circuits can improve performance of our method by reducing the number of gates in adiabatic time evolution.

Now, we move on to the three-site periodic system. We start with all spins up. The time evolution circuit for the XY part is implemented pairwise using the XY part of two qubit circuit for each Trotter step. The data obtained from the quantum computer with readout error correction is shown in Fig.~\ref{fig:3 site Exp data} along with the scaled data so as to match the end points from the simulator data. The extrapolations are shown in Fig.~\ref{fig:3 site Extraploation to find transition}. A quartic fit to the exact values gives the critical $B_z$ value to be 0.9987J. A quartic fit to the simulator values gives the critical $B_z$ to be 0.982J. The linear fit to the crossings from the scaled data of IBM quantum computer give the phase transition point to be 1.11J for IBM Yorktown, 0.97J for IBM Montreal and 0.99J for the IBM Casablanca machines. The actual transition is at 1.0J. These fitted values are reasonably close to the actual value.

\section{Conclusion}

In this work, we propose a method for finding zero-temperature phase diagrams that is robust and can be carried out on quantum computers. The approach requires us to introduce a symmetry breaking term into the Hamiltonian, determine approximate phase diagrams for the symmetry-broken system, and then extrapolate to the limit where the symmetry-breaking field vanishes. To verify that this approach works, we have worked out practical details for how to run these circuits on a quantum computer when the number of spins is 2 or 3. The results from the quantum computers agree well with the exact results and are able to predict the phase boundaries within a few percent. This illustrates that the approach used here, based on adiabatic state preparation, can work on NISQ machines and has the potential to be able to be applied to larger systems, even ones where we do not know the phase diagram \textit{a priori}.

Note that the case we examined here, the XY model in a $z$-oriented magnetic field, is probably the most difficult problem to examine, because the number of level crossings increases with the system size. For most quantum phase transitions between different symmetry states, the number of phase boundaries should depend only weakly on the system size.

 In order to show that this approach also applies to larger systems, we simulate the magnetization for a ten site system in Fig.~\ref{fig:10site curve} using a similar local adiabatic time evolution for 1000 Trotter steps for $\gamma$ = 50. Extracting the phase transitions using our methodology works well for such a system, as can be seen by comparing the two lines in the figure.

\section*{Acknowledgements}
For this work, the planning, formal development, and circuit development (A.F., J.K.F, and A.F.K.), as well as part of the manuscript writing (J.K.F. and A.F.K.) was supported by the Department of Energy, Office of Basic Energy Sciences, Division of Materials Sciences and Engineering under Grant No. DE-SC0019469.
The circuit execution and manuscript writing (A.F.) was supported by the National Science Foundation under Grant No. DMR-1752713.
E.Z. was supported by the U.S. Department of Energy, Office of Science, Office of Advanced Scientific Computing Research (ASCR), Quantum Computing Application Teams (QCATS) program, under field work proposal number ERKJ347 for her work in the summer of 2020 in developing the formalism and exact diagonalization codes for this project.
J. K. F. was also supported by the McDevitt bequest at Georgetown.
We acknowledge fruitful discussions with Brian Rost.
We acknowledge the use of IBMQ via the IBM Q Hub at NC State for this work. The views expressed are those of the authors and do not reflect the official policy or position of the IBM Q Hub at NC State, IBM or the IBM Q team. We acknowledge the use of Qiskit software package~\citep{Qiskit} for doing the quantum simulations.
The data for the figures are available at \url{https://doi.org/10.5061/dryad.z8w9ghxdq}

\bibliography{xyadiabatic}


\end{document}